\documentclass[a4paper]{jpconf}
\usepackage{graphicx,psfrag,color}
\usepackage{mathbbol,amsmath,amsfonts,amssymb,bm,sidecap,hyperref,dsfont}
\def\im{{\sf {i}}}

\begin{document}

\title{What is a particle-conserving Topological Superfluid? 
{\large The fate of Majorana modes beyond mean-field theory}}

\author{Gerardo Ortiz$^1$ and Emilio Cobanera$^2$}

\address{$^1$ Department of Physics, Indiana University, Bloomington,
IN 47405, USA}
\address{$^2$ Department of Physics and Astronomy, Dartmouth
College, 6127 Wilder Laboratory, Hanover, NH 03755, USA}

\ead{ortizg@indiana.edu}

\begin{abstract}
We establish a criterion for characterizing superfluidity in 
interacting, particle-number conserving systems of fermions as 
topologically trivial or non-trivial. Because our criterion is 
based on the concept of many-body fermionic parity switches, it
is directly associated to the observation of the fractional 
Josephson effect and indicates the emergence of zero-energy modes 
that anticommute with fermionic parity. We tested these ideas on 
the Richardson-Gaudin-Kitaev chain, a particle-number conserving 
system that is solvable by way of the algebraic Bethe ansatz, and 
reduces to a long-range 
Kitaev chain in the mean-field approximation. Guided by its closed-form 
solution, we introduce a procedure for constructing 
many-body Majorana zero-energy modes of gapped  
topological superfluids in terms of coherent superpositions of 
states with different number of fermions. 
We discuss their significance and the physical conditions  required 
to enable quantum control in the light of superselection rules. 
\end{abstract}

\section{Introduction}

Most investigations of fermionic condensed matter are based on a type
of mean-field approximation popularized by Bogoliubov \cite{MF}. 
One recent and conspicuous example is the ten-fold way, a topological
classification of fermionic systems based on three discrete 
--- time reversal, charge conjugation, and chiral --- symmetries and 
K-homology \cite{Ryu2010,periodic_table}. The main reasons for the
prevalence of the mean-field approximation are clear. First, it leads 
to a very intuitive and natural picture of fermionic systems thanks to 
Landau's theory of Fermi liquids and their quasi-particles. Second, 
it leads to a mathematically simple Lie-algebraic formalism by which
the problem of diagonalizing the Hamiltonian in Fock space becomes
of polynomial complexity in the total number of degrees of freedom. 
And last but not least, topological invariants such as (full or partial) 
Chern numbers, Berry phases \cite{TScs}, Bott and Hopf indexes, 
and others are easily associated to mean-field theories and evaluated 
for concrete instances. 

The thermodynamic state of a fermionic superfluid (or superconductor) is characterized by 
the {\it spontaneous} breaking of the global continuous $U(1$) symmetry related to particle-number 
conservation. 
The positive features of the mean-field approximation for
fermions come at a surprising cost in the context of superfluidity
(electrically neutral fermions) or superconductivity (charged fermions): 
the {\it explicit} breaking of the symmetry of particle-number conservation.
However, electronic matter is composed of interacting 
electrons whose number is locally conserved. Whether this
mismatch between models and systems being modeled matters 
or not is bound to depend on the physical quantities to be computed.  
Many calculations of thermodynamic and transport properties 
of fermionic superfluids have firmly established the 
phenomenological success of particle-number non-conserving 
mean-field theories. But these successes do not imply that 
every experimentally accessible feature of the fermionic 
superfluid state is well described by breaking particle-number 
conservation, see for example Ref.\,\cite{delft01}. 

In this paper we will set up the ground for a systematic 
investigation of the interplay between the mean-field topological 
classification of superfluid systems of fermions and more 
realistic models where, for closed systems, the number of 
fermions is conserved. Our discussion will be organized around 
three fundamental and interrelated questions. 
\begin{itemize}
\item
What makes a particle-number conserving fermionic superfluid/superconductor
topologically trivial or non-trivial?
\item
What are the experimental signatures of fermionic topological superfluidity ? 
\item
What is the fate of Majorana zero-energy modes beyond
the mean-field approximation? And even more basically, what is their
very meaning?
\end{itemize}

Concern with the role of particle (non-)conservation in the 
mean-field theory of superconductivity is as old as the theory 
itself \cite{footnote1}. We feel prompted to revisit this issue 
by the recent {\it experimental} efforts to detect and control 
Majorana zero-energy modes. The presence of Majorana modes has 
been typically considered a key mean-field manifestation of topological fermion 
superfluidity since the work of Ref.\,\cite{green_read}.
These quasi-particles emerge from the interplay  
between the existence of a topologically non-trivial  vacuum and 
a, typically, symmetry-protected physical boundary (or defect). 
In recent literature, this connection goes under the name of 
bulk-boundary correspondence. Because of the expected resilience 
against decoherence and non-Abelian braiding properties, 
Majorana modes, or simply Majorana fermions for short,
are key components of many blueprints of quantum-information processing 
devices. Given that either electrons or fermionic atoms are in fact 
locally conserved, it is imperative to investigate 
the conditions for the emergence of Majorana fermions 
and procedures for their experimental detection beyond mean field.

Our work here suggests the idea that a Majorana fermion may be quantum 
controlled is so deeply rooted in the mean-field picture that it 
might not have a natural counterpart in more realistic particle-number
conserving frameworks. Briefly stated, if a zero-energy 
mode of a superfluid system creates a superposition of states that 
differ in particle number/electric charge, it may not be possible 
to manipulate this mode without exchanging particles with an environment, 
as opposed to exchanging, say, energy only. Then the question becomes 
whether it is possible in practice to exchange {\it coherently} (charged) 
particles with a reservoir big enough to grant the mean-field picture 
of the (sub)system of interest. This claim is ultimately rooted in
the venerable \cite{wightman95}, but still much investigated and 
debated \cite{bartlett07}, subject of superselection rules.
 
\begin{figure}[tb]
\centerline{\includegraphics[width=0.6\columnwidth]{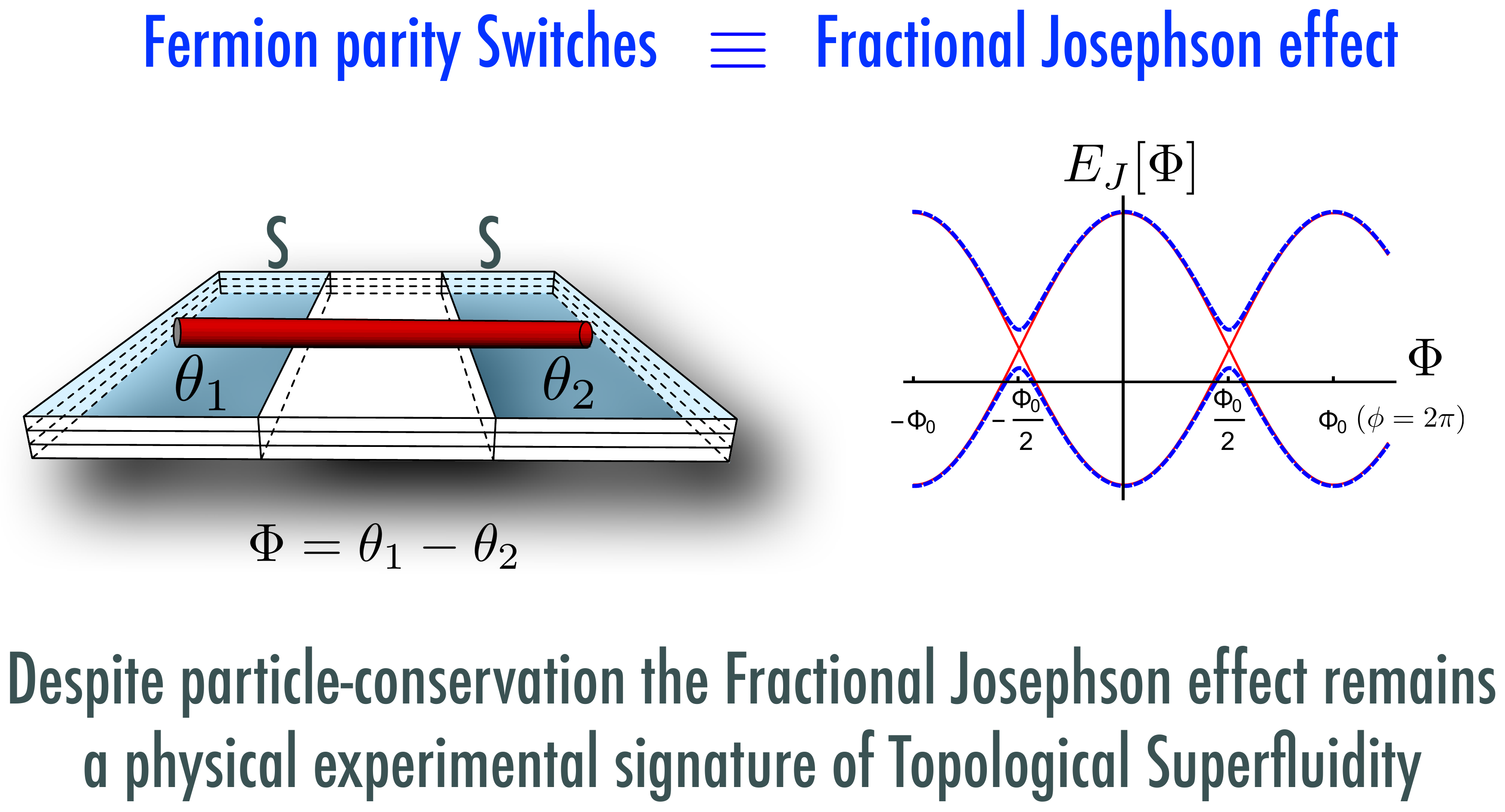}}
\caption{(Left panel) Schematics of an SNS Josephson junction. (Right panel) Standard 
($2\pi$-periodic)
and fractional ($4\pi$-periodic) Josephson effects depicting the Josephson energy 
$E_J[\Phi]$ as a function of the magnetic flux $\Phi=\frac{\phi}{2\pi}\Phi_0$, where 
 $\Phi_0$ is the superconducting flux quantum.}
\label{fig0}
\end{figure}

Our last statement requires some clarification, since it 
takes for granted the existence of Majorana modes for closed, hence 
necessarily particle-number conserving, systems. Here we will 
investigate whether this assumption is reasonable based on our work
published in Ref.\;\cite{rgkchain}. By exploiting the algebraic Bethe ansatz,
we succeeded in establishing and characterizing topological
superconductivity for prototypical particle number-conserving, and 
thus necessarily interacting, superconducting chains beyond mean-field 
theory. Before this work, it was not known how to test for topological
fermion superfluidity in number-conserving systems. 
The fact that a witness based on fermionic parity switches works 
in spite of the conservation of particle number can be firmly established 
thanks to the realization of an exactly-solvable topological fermion 
superfluid, the Richardson-Gaudin-Kitaev ({\sf RGK}) wire. Moreover, 
the RGK wire allows derivation of an exact topological invariant. 
Consequently, it follows that the fractional Josephson  effect remains 
a signature of topological fermion superfluidity (see Fig. \ref{fig0}).


To summarize, in this paper we intend to convey a certain amount 
of caution as to what is physically possible in terms of manipulation 
and control of Majorana modes, and hope to shed some light on this 
exciting field \cite{Leggett}. The outline is as follows.   
We begin in Section \ref{toy} by presenting an exactly solvable model 
of a particle-conserving superfluid that, as will be shown explicitly,  
displays topologically trivial and non-trivial phases \cite{rgkchain}. 
In Section \ref{charact} we consider the problem of characterizing a 
particle-conserving fermionic superfluid as topologically trivial or 
non-trivial. This problem has been shown to have a definite answer 
only very recently \cite{rgkchain}. As explained in this paper, a topological 
fermion superfluid  (or superconductor), 
in addition to the global $U(1)$ symmetry of particle-number conservation,  
{\it spontaneously}  breaks  the discrete $\mathbb{Z}_2$ 
symmetry of fermionic parity. In Section \ref{NCMajoranas} we 
focus on the meaning and nature of Majorana zero-energy modes in gapped,
interacting many-electron systems.  (One may compare to 
recent work based on a quasi-exactly solvable, particle-number conserving,
two-leg ladder model of spinless fermions with open boundary conditions  
and a gapless  excitation spectrum \cite{Iemini,Buchler}. The concocted  
Majorana-like modes are not related to total fermionic parity.) And in Section 
\ref{SSR} we discuss the physical picture that emerges, from the standpoint 
of superselection rules, as far as the quantum manipulation and control of
Majorana modes is concerned. We conclude in Section \ref{outlook} with an 
outlook.

\section{Particle-number Conserving Fermionic Superfluids: The {\sf RGK} chain} 
\label{toy}

We now present a model, dubbed the {\sf RGK}
chain, introduced and solved in Ref. \cite{rgkchain}. The {\sf RGK} chain 
is the first example of an interacting, particle-conserving, fermionic superfluid 
in one spatial dimension shown to display 
a topologically non-trivial superfluid phase. It was designed to benchmark 
possible criteria of topological superfluidity in number conserving systems. 

\subsection{The Hamiltonian in position and momentum representations}
The Hamiltonian of the {\sf RGK} chain, in the momentum representation, is  
given by
\begin{eqnarray}
H_{\sf RGK}=\!\!\!\sum_{k\in  {\cal S}_{k}^{\phi}}
\varepsilon_k \, \hat{c}^\dagger_k\hat{c}_k^{\;}
- 8 G\hspace*{-0.3cm}\sum_{k,k' \in {\cal S}_{k+}^{\phi}}\hspace*{-0.25cm}
\eta_k \eta_{k'}
\hat{c}_{k}^\dagger \hat{c}_{-k}^\dagger \hat{c}_{-k'}^{\;}\hat{c}_{k'}^{\;} ,
\label{HamiltRGnew}
\end{eqnarray}
in terms of spinless (or fully spin-polarized) fermion creation 
operators $ \hat{c}^\dagger_k$, with momentum $k$-dependent single-particle
spectrum 
\begin{eqnarray}
\varepsilon_k=-2t_1\cos k-2t_2 \cos 2k ,
\end{eqnarray}
where $t_1, t_2$ are the nearest and next-nearest neighbor hopping amplitudes, 
and $G>0$ the attractive interaction strength. The interaction
strength is modulated by the potential
\begin{eqnarray}
\eta_k =\sin (k/2)\sqrt{t_1 + 4 t_2 \cos^2 (k/2)}\label{etakdef}
\end{eqnarray}
odd in \(k\), $\eta_k=-\eta_{-k}$, as is characteristic of
$p$-wave superconductivity. 

The pair potential and the  single-particle spectrum
are connected by the simple relation
\begin{eqnarray}
4\eta_k^2=\varepsilon_k+2 t_+ \quad (t_+=t_1+t_2). 
\end{eqnarray}
This property of the model is the key for achieving
exact solvability. Nonetheless,  their functional forms have 
been chosen so that they also realize a new exactly solvable 
model that is physically sound in position, as well as momentum, space.
\begin{figure}[tb]
\centerline{\includegraphics[width=0.4\columnwidth]{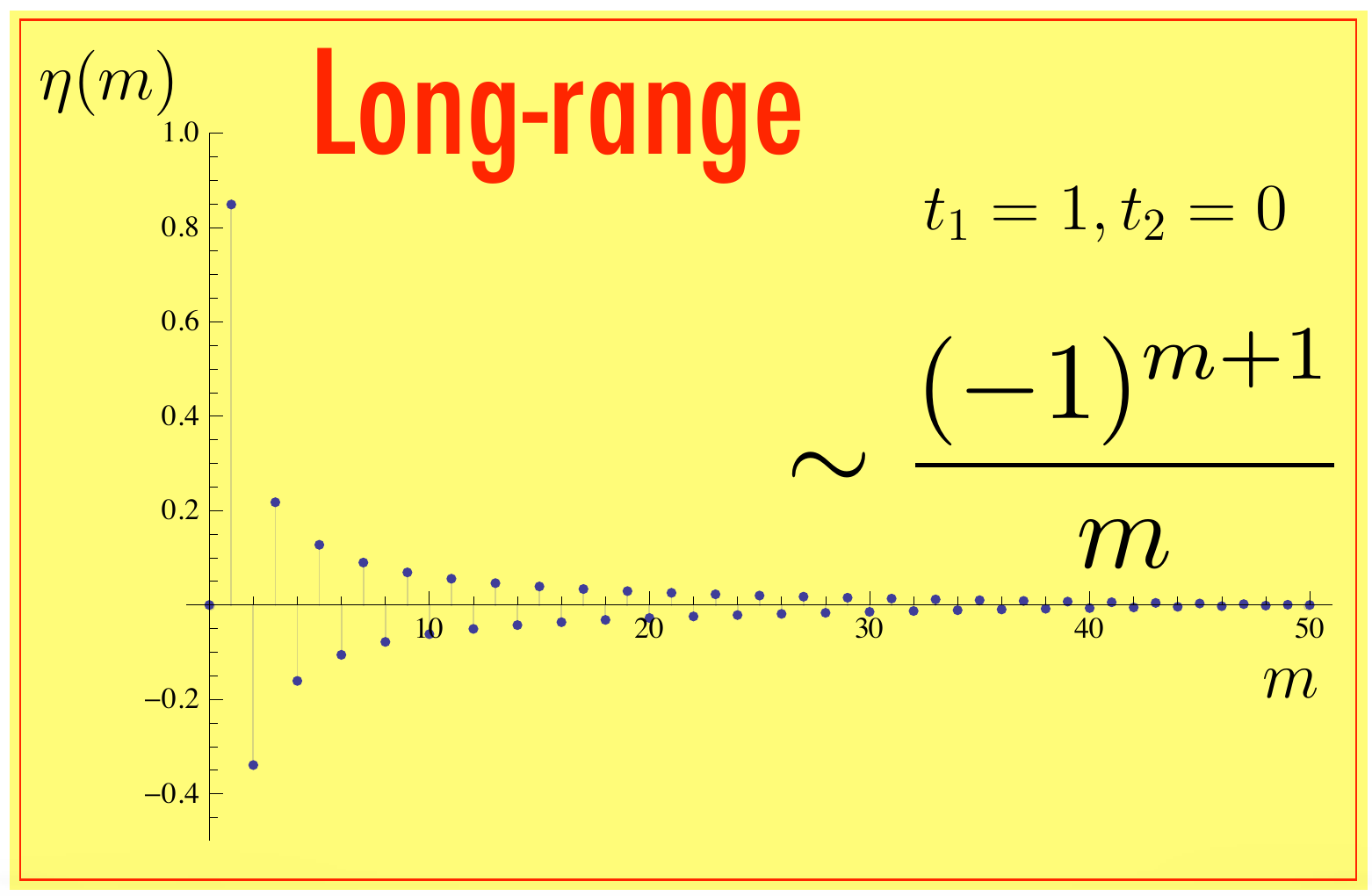}}
 \caption{Functional form of the pairing interaction $\eta(m)$ for $L=100$,  
 $t_1=1$ and $t_2=0$.}
 \label{fig1}
\end{figure}
In position representation, we define 
$c_j=L^{-1/2}\sum_{k \in {\cal S}_{k}^{\phi}} e^{\im j  k}\, 
\hat{c}_{k}$ for a chain of length $L$, measured in units of the 
lattice constant. We take $\phi$-dependent boundary conditions 
$c_{j+L}=e^{\im \phi/2}c_j$. In a ring geometry, periodic boundary 
conditions ($\phi=0$) correspond to enclosed flux $\Phi=0$ and 
antiperiodic boundary conditions ($\phi=2\pi$) correspond to 
$\Phi=\Phi_0=h/2e$. The resulting sets of allowed momenta 
\({\cal S}_{k}^{\phi}\) are   
\begin{eqnarray}
{\cal S}_{k}^{0}={\cal S}_{k+}^{0}\oplus {\cal S}_{k-}^{0}\oplus \{0,-\pi\} \ \ 
\mbox{ and } \  \
{\cal S}_{k}^{2\pi}={\cal S}_{k+}^{2\pi}\oplus {\cal S}_{k-}^{2\pi},
\end{eqnarray}
with 
\begin{eqnarray}
{\cal S}_{k\pm}^{0}=L^{-1}\{\pm 2\pi, \pm 4\pi,
\cdots, \pm(\pi L-2\pi)\} \ \  
\mbox{ and  } \ \ {\cal S}_{k\pm}^{2\pi}=L^{-1}\{\pm\pi, \pm 3\pi,
\cdots, \pm(\pi L -\pi)\}  . 
\end{eqnarray}
The number of momenta per sector is 
Card$[{\cal S}_{k\pm}^{\phi}]=\frac{L}{2}-\delta_{\phi,0}$, 
so that it totals to Card$[{\cal S}_{k}^{\phi}]=L$. After Fourier transformation, 
the {\sf RGK} Hamiltonian in the position representation is given by
\begin{eqnarray}
H_{\sf RGK}&=&-\sum_{i=1}^L \sum_{r=1}^2\left(t_r \, c_{i}^{\dagger}c_{i+r}^{\;}
+{\rm H.c.}\right) - \ 2G \, I^\dagger_\phi I^{\;}_\phi, \ \ \mbox{    where }\\
I_\phi&\equiv& 2 \im \sum_{k \in {\cal S}_{k+}^{\phi}}
\eta_k \ \hat{c}_k\hat{c}_{-k}=\sum_{i>j}^L  \eta(i-j) \, c_{i}^{\;}c_{j}^{\;}.
\end{eqnarray}

There are at least two cases in which the pairing function 
$\eta(m)$ can be determined in closed form by Fourier transformation 
of Eq.\ \eqref{etakdef}. For $t_1=0$ and $t_2\neq 0$,
$\eta(m)=\sqrt{t_2} \, \delta_{m1}$, and so we obtain nearest-neighbor 
pairing only. For $t_1\neq 0$ and $t_2= 0$ instead we obtain 
\begin{eqnarray}
\eta(m)=\frac{(-1)^m \, 8 \sqrt{t_1}}{\pi}\frac{m}{1-4m^2} ,
\;\;{\rm for}\;\;L\rightarrow\infty,
\label{etamdef}
\end{eqnarray}
describing a long-range pairing interaction with a slow $1/m$ decay 
with distance $m=i-j$, see Fig. \ref{fig1}. 
In general $\eta(m)$ is a monotonically decaying function of $m$ with
$\eta(0)=0=\eta(L/2)$, $\eta(m)>0$ or $<0$ for $m$ odd or even, respectively.

This long-ranged pairing interaction, a main difference with the 
original Kitaev chain \cite{Kitaev2001}, allows for an exact solution 
beyond the mean-field approximation and a gapped spectrum in spite
of the one-dimensional character of the model. As we shall see in a 
moment, the long-range coupling allows as well for a topologically 
non-trivial phase. It may also be physically relevant for chains of 
magnetic nanoparticles on a superconducting substrate \cite{Cho11,Nad13},
which have recently been shown to support topologically protected Majorana 
zero-modes in the presence of a long-range coupling \cite{Pie13}. 

\subsection{Mean-field approximation}

Before we present the exact solution of the {\sf RGK} chain, we would 
like to establish whether it displays a non-trivial topological 
phase in the mean-field approximation. For simplicity we 
consider the pairing function \eqref{etamdef}, that is, \(t_2=0\). 
The mean-field approximation  is obtained from the substitution 
\begin{eqnarray}
2GI^\dagger_\phi I^{\;}_\phi\rightarrow \Delta^* I^{\;}_\phi +\Delta \,  
I^\dagger_\phi,
\end{eqnarray}
with gap function 
\begin{eqnarray}
\Delta=2G\langle I^{\;}_{\phi}\rangle=e^{\im \theta} |\Delta|.
\end{eqnarray}

Let us define Majorana fermion operators
$a_i=e^{-\im \theta/2}c^{\;}_i + e^{\im \theta/2}c^\dagger_i$,
$\im b_i=e^{-\im \theta/2}c^{\;}_i - e^{\im \theta/2}c^\dagger_i$.
In terms of these degrees of freedom the mean-field Hamiltonian is
\begin{equation}
H_{\sf mf}=
\frac{\im t_1}{2}\sum_{i=1}^{L-1}(b_ia_{i+1}-a_ib_{i+1})-
\frac{\im}{2} \sum_{i>j}^{L}\Delta_{i-j}(b_ia_{j}+a_ib_{j}),
\end{equation}
with $\Delta_{m}=|\Delta| \, \eta(m)$, and can be shown to display 
a topological phase characterized by power-law Majorana edge modes 
and the associated \(4\pi\)-periodic Josephson effect. Due to the 
long-range nature of the interaction, the wavefunction of the 
Majorana edge modes decays algebraically rather than exponentially in 
the bulk, and their energy approaches zero as a power law in $1/L$.

\subsection{Exact solution} 

In Ref. \cite{rgkchain} we showed that the {\sf RGK} chain is 
exactly solvable. To this end we re-wrote it in the algebraic 
form
\begin{align} \!\!\!\!\!\!
H_{\sf RGK}={}& 8 H_{\phi} +\delta_{\phi,0} \, (\varepsilon_0 \hat{c}^\dagger_0 \hat{c}_{0}^{\;}+
\varepsilon_{-\pi} \hat{c}^\dagger_{-\pi}\hat{c}_{-\pi}^{\;}) -4 t_+  \, S^z+C_\phi , \ \mbox{ with }\\
H_\phi={}&\sum_{k \in {\cal S}_{k+}^{\phi}}\eta_{k}^2 \ S_{k}^{z}-G \!\!\!\!\sum_{k,k' \in
{\cal S}_{k+}^{\phi} } \!\!\! \eta_{k}\eta_{k^{\prime}
} \ S_{k}^{+}S_{k^{\prime}}^{-},
\label{HamiltRG0}
\end{align}
$C_\phi= 2 t_2 \delta_{\phi,0}$, 
$S_{k}^{z}=\frac{1}{2}(  \hat{c}^\dagger_k \hat{c}_{k}^{\;}
+\hat{c}^\dagger_{-k}\hat{c}_{-k}^{\;}-1)$,  and
$S_{k}^{+}=\hat{c}_{k}^\dagger \hat{c}_{-k}^\dagger$
for each pair \((k,-k)\) of pairing-active momenta. The operators $S_{k}^{z}, S_{k}^{\pm}$
satisfy the algebra of
\({\rm SU}(2)\). It follows that 
\begin{eqnarray}
S^z= \sum_{k \in {\cal S}_{k+}^{\phi}} S_{k}^{z}
\end{eqnarray}
represents a conserved quantity, $[H_\phi,S^z]=0=[H_{\sf RGK},S^z]$.
The relation to the total fermion number operator \(\hat{N}\) is
\begin{eqnarray}
 2S^z=\left\{
\begin{array}{lcl}
\hat{N}-
(\hat{c}^\dagger_0 \hat{c}_{0}^{\;}+
\hat{c}^\dagger_{-\pi}\hat{c}_{-\pi}^{\;})-\Big(\frac{L}{2}-1\Big)&
\mbox{if} & \ \ \ \phi=0\\
\hat{N}-\frac{L}{2}& \mbox{if}& \ \ \ \phi=2\pi
\end{array}\right. .
\end{eqnarray}
\begin{figure}[htb]
\centerline{\includegraphics[width=0.65\columnwidth]{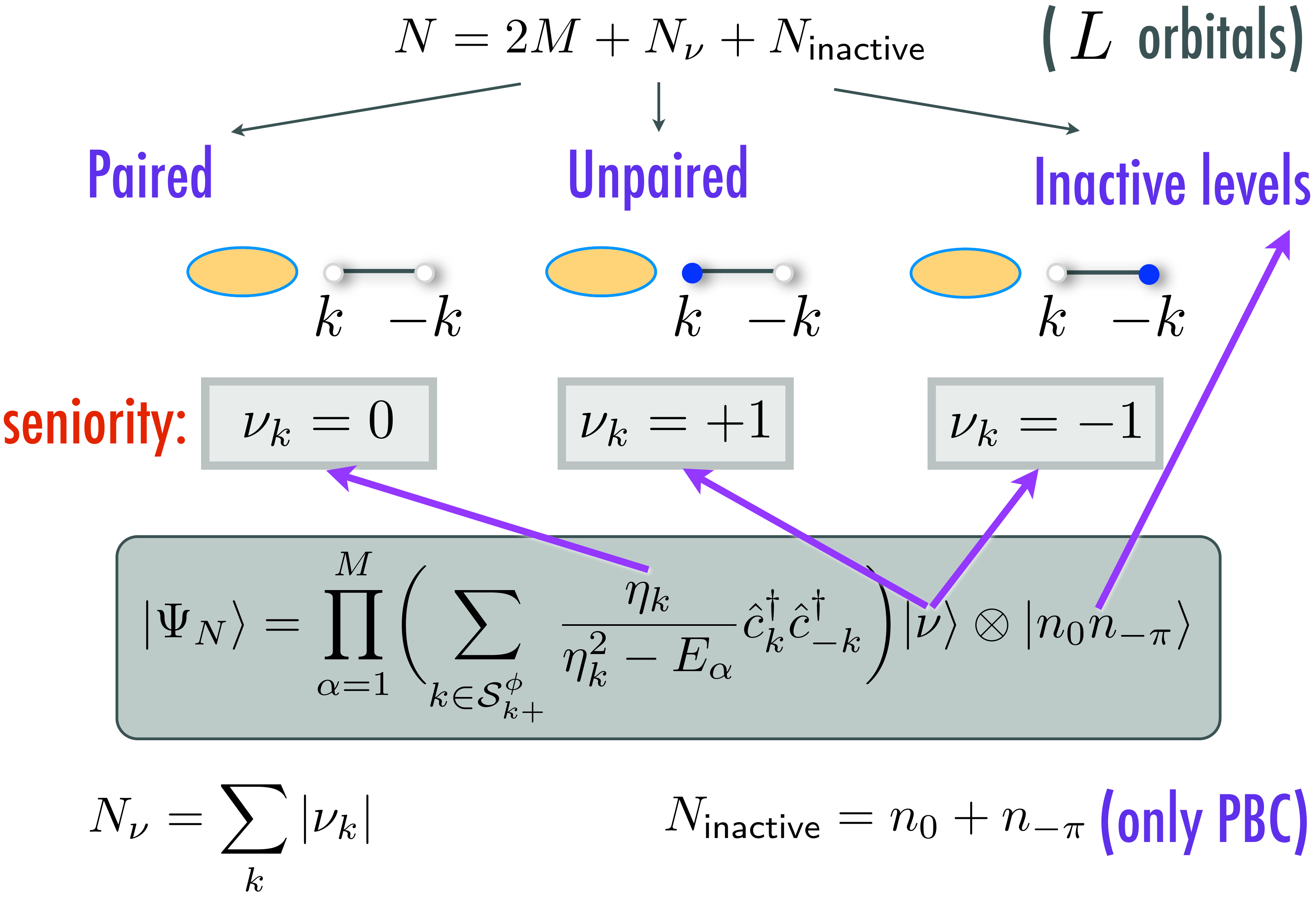}}
 \caption{The eigenstates $| \Psi_N\rangle$ of $H_{\sf RGK}$ are constructed out of 
 three different kinds of electrons ($s_k=\frac{1-|\nu_k|}{2}$): paired, unpaired, and inactive (only 
 for periodic boundary conditions ({\sf PBC})).}
 \label{fig2}
\end{figure}

Written in this form, one immediately recognizes $H_\phi$ as 
an exactly solvable pairing Hamiltonian belonging to the hyperbolic 
family of Richardson-Gaudin integrable models \cite{Duk04,ortiz2005}.  
Eigenstates with exactly $2M+N_\nu$ fermions are given by
\begin{eqnarray}
|\Phi_{M,\nu}\rangle =\prod_{\alpha=1}^M \biggl( \sum_{k \in {\cal S}_{k+}^{\phi}}
\frac{\eta_k}{\eta_k^2-E_\alpha} \hat{c}_{k}^\dagger \hat{c}_{-k}^\dagger \biggr)
| \nu\rangle ,
\label{wavefunction}
\end{eqnarray}
where $M$ is the number of fermion pairs. 
The state $|\nu\rangle$ with $N_\nu$ unpaired fermions satisfies
$S^-_k| \nu\rangle=0$ for all $k$. Moreover, 
$S^z_k| \nu\rangle=-s_k | \nu\rangle$, with $s_k=0$ if the level 
$k$ is singly-occupied or $s_k=1/2$ if it is empty (see Fig. \ref{fig2}). The corresponding energy levels are
$E_{M,\nu}=\langle \nu | H_\phi|\nu \rangle+\sum_{\alpha=1}^M E_\alpha$, 
with spectral parameters $E_\alpha$ determined by the Richardson-Gaudin 
(Bethe) equations
\begin{eqnarray}
\sum_{k \in {\cal S}_{k+}^{\phi}}
\frac{s_k}{\eta_k^2-E_{\alpha}}-\sum_{\beta\left(
\neq\alpha\right)  }\frac{1}{E_{\beta}-E_{\alpha}}=\frac{Q_\phi}{E_{\alpha}} ,
\label{Betheeq}
\end{eqnarray}
where $Q_\phi=1/2G-\sum_{k \in {\cal S}_{k+}^{\phi}}s_k+M-1$.

For periodic boundary conditions ($\phi=0$), the two momenta $k=0,-\pi$
are not affected by the interactions and must be included separately.
Then, eigenvectors of $H_{\sf RGK}$ are given by
\begin{eqnarray}
| \Psi_N\rangle= |\Phi_{M, \nu}\rangle \otimes |n_0n_{-\pi}\rangle,
\end{eqnarray}
where $N=2M+N_\nu+n_0+n_{-\pi}$ is 
the total number of fermions and $n_{0},n_{-\pi}\in\{0,1\}$. See Fig. \ref{fig2}.

\subsection{Quantum phase diagram} 

The quantum phase diagram of the {\sf RGK} chain is
determined from the analytical dependence of its ground 
energy ${\cal E}_0(\rho,g)$ on the density $\rho=N/L$ and scaled coupling
strength $g=GL/2$. Depending on the boundary condition $\phi$ and fermion-number parity, 
one should consider either $N_\nu=0$ or 1. For periodic boundary 
conditions, since the levels $k=0,-\pi$ decouple from the rest, 
$N_\nu=0$ for both even and odd $N$. If $N$ is odd, then the 
unpaired particle occupies the $k=0$ level without blocking an active
level. For antiperiodic boundary conditions the ground state
has $N_\nu=0$ for $N$ even, while for $N$ odd it has $N_\nu=1$ 
with blocked level $k_0$. The resulting ground state energy is 
given by
\begin{equation}
{\cal E}^{\phi}_0(N)= 8\sum_{\alpha=1}^M \! E_\alpha- 4 t_+  
\,M + J_{\phi,0} + \delta_{N_\nu,1} (4 \eta_{k_0}^2-2t_+),
\end{equation}
where $J_{\phi,0}=\delta_{\phi,0} \, (\varepsilon_0  \, \delta_{n_0,1}+
\varepsilon_{-\pi}\, \delta_{n_{-\pi},1})$. In the thermodynamic limit 
($N,L \rightarrow \infty$, such that $\rho$ is kept constant) the energy 
density becomes
\begin{equation}
e_0\equiv\lim_{L\rightarrow\infty}{\cal E}^\phi_0/L=-2t_+
\rho-\frac{4}{g}\Delta^2+\frac{4}{\pi}\int^{\pi}_0 \eta^2_k v^2_k \, dk ,
\end{equation}
where the chemical potential $\mu$ and gap function $\Delta$ are related 
to the original parameters of the model by
\begin{align}
&\frac{2\pi}{g}=\int_{0}^{\pi}\frac{\eta_{k}^{2}}{E_{k}} \, dk,\;\;
\rho=\frac{1}{\pi}\int_{0}^{\pi}v_{k}^{2} \, dk,\\
&E_k=\sqrt{\left(\frac{\eta^2_k}{2}-\mu\right)^2+\eta^2_k \Delta^2},\;\;
v^2_k=\frac{1}{2}-\frac{\eta^2_k-2\mu}{4E_k} .
\end{align}

\begin{figure}[tb]
\centerline{\includegraphics[width=0.55\columnwidth]{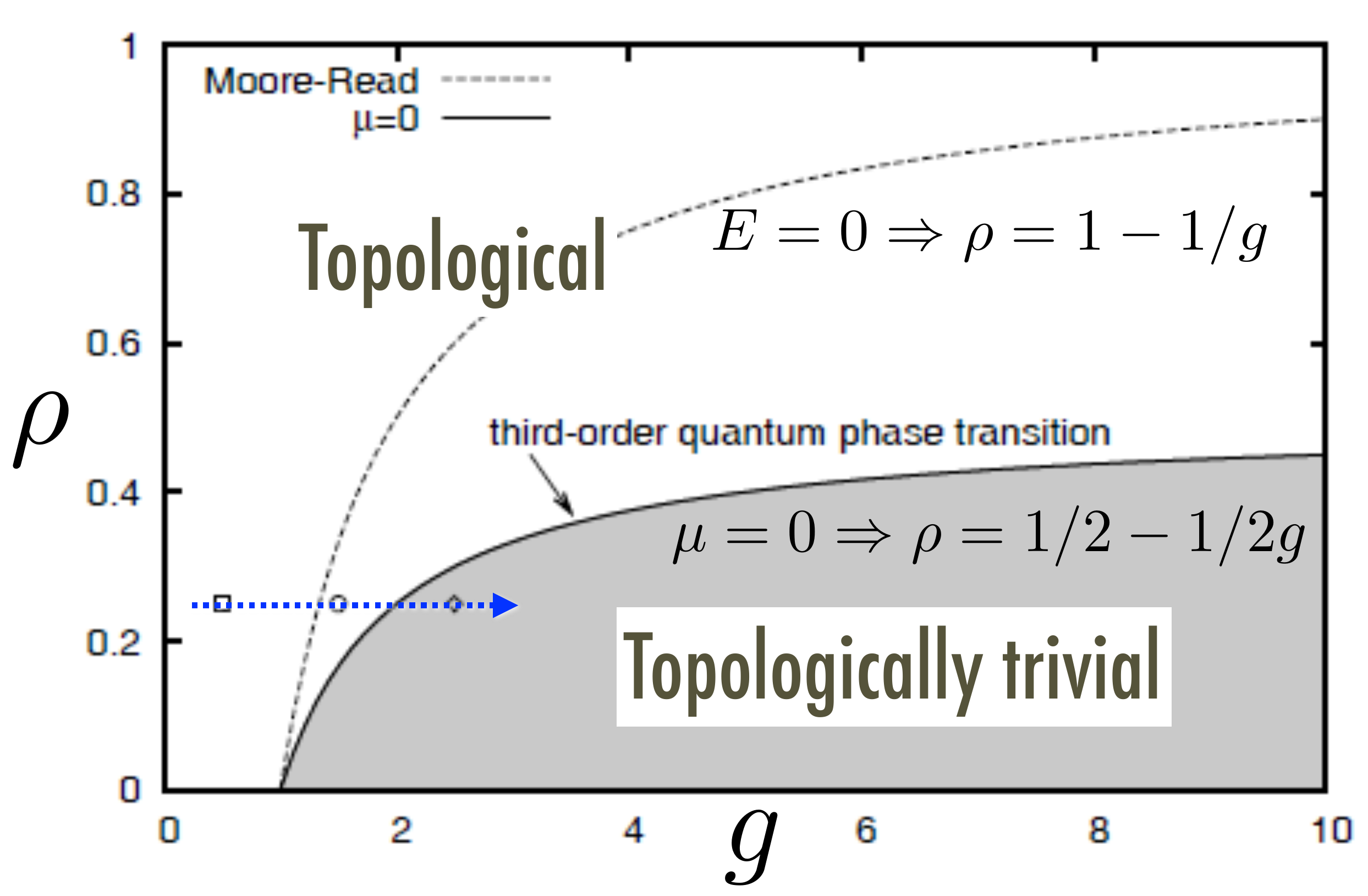}}
 \caption{Quantum phase diagram of the {\sf RGK} 
wire in the $(\rho,g)$-plane. Dashed and full lines represent 
the Moore-Read ($g^{-1}=1-\rho$) and Read-Green ($g^{-1}=1-2\rho$) 
boundaries, respectively.  As shown in the text, the weak-pairing phase is topologically non-trivial 
while the strong pairing phase is topologically trivial. The horizontal dashed arrow corresponds to 
the density $\rho=1/4$. }
 \label{fig3}
\end{figure}

The resulting phase diagram is shown in Fig.\ \ref{fig3}.
The {\sf RGK} chain is gapped for all $g>0$, except for the Read-Green 
coupling $g=g_c=1/(1-2\rho)$ where
it becomes critical in the thermodynamic limit, independently of
the choice of boundary conditions $\phi$. This critical 
line defines the phase boundary separating 
weak from strong pairing phases, and thus is a line of non-analyticities. 
At \(g_c\) a cusp develops in the second derivative of $e_0$, that leads 
to singular discontinuous behavior of the third-order derivative. Hence 
the transition from a weakly-paired to a strongly-paired fermionic superfluid 
is of third order, just like for the two-dimensional chiral $p$-wave 
superconductor \cite{rombouts2010,Lerma2011}. Which one of these
two superfluid phases, if any, may be properly characterized as topologically 
non-trivial?

\section{Fermionic Parity Switches and the Fractional Josephson Effect}
\label{charact}

 Reference \cite{rgkchain} introduced a 
quantitative criterion
for establishing the emergence of topological superfluidity in particle-number 
conserving, many-fermion systems. The criterion exploits the behavior of 
the ground state energy of a system of $N$, and $N\pm1$ particles,
for both periodic and antiperiodic boundary conditions. The emergence 
of topological order in a superconducting wire, closed in a ring and 
described in mean-field, is associated with switches in the ground-state 
fermion parity ${\cal P}(\phi)$ upon increasing the enclosed flux 
$\Phi=(\phi/2\pi)\times \Phi_0$  
\cite{Kitaev2001,Kes13,Bee13,Sau13,Hai14,Cre14,Bee14}. Any spin-active
superconductor, topologically trivial or not, may experience a crossing 
of the ground state energies for even and odd number of electrons 
\cite{Sak70,Bal06,Cha12,Lee14}. Regardless of spin, what matters is the 
number of crossings $N_{X}$ between $\Phi=0=\phi$ and $\Phi=\Phi_0$, 
$\phi=2\pi$. The superconductor is topologically non-trivial if $N_{X}$ is 
odd, otherwise it is trivial.

In the many-body, number conserving, case we need to identify the relevant 
parity switches signaling the emergence of a topological fermion superfluid 
phase. Our exact solution gives us access to ${\cal P}(\phi)$ only at 
$\phi=0$ and $\phi=2\pi$, but this is sufficient to determine whether 
$N_X$ is even or odd. Notice that odd $N_X$ means that the flux $\Phi$ 
should be advanced by $2\Phi_0$, rather than $\Phi_0$, in order to return 
to the initial ground state. This is the essence of the $4\pi$-periodic 
Josephson effect \cite{Kitaev2001,Sen01}.

\begin{figure}[htb]
\centerline{\includegraphics[width=0.65\columnwidth]{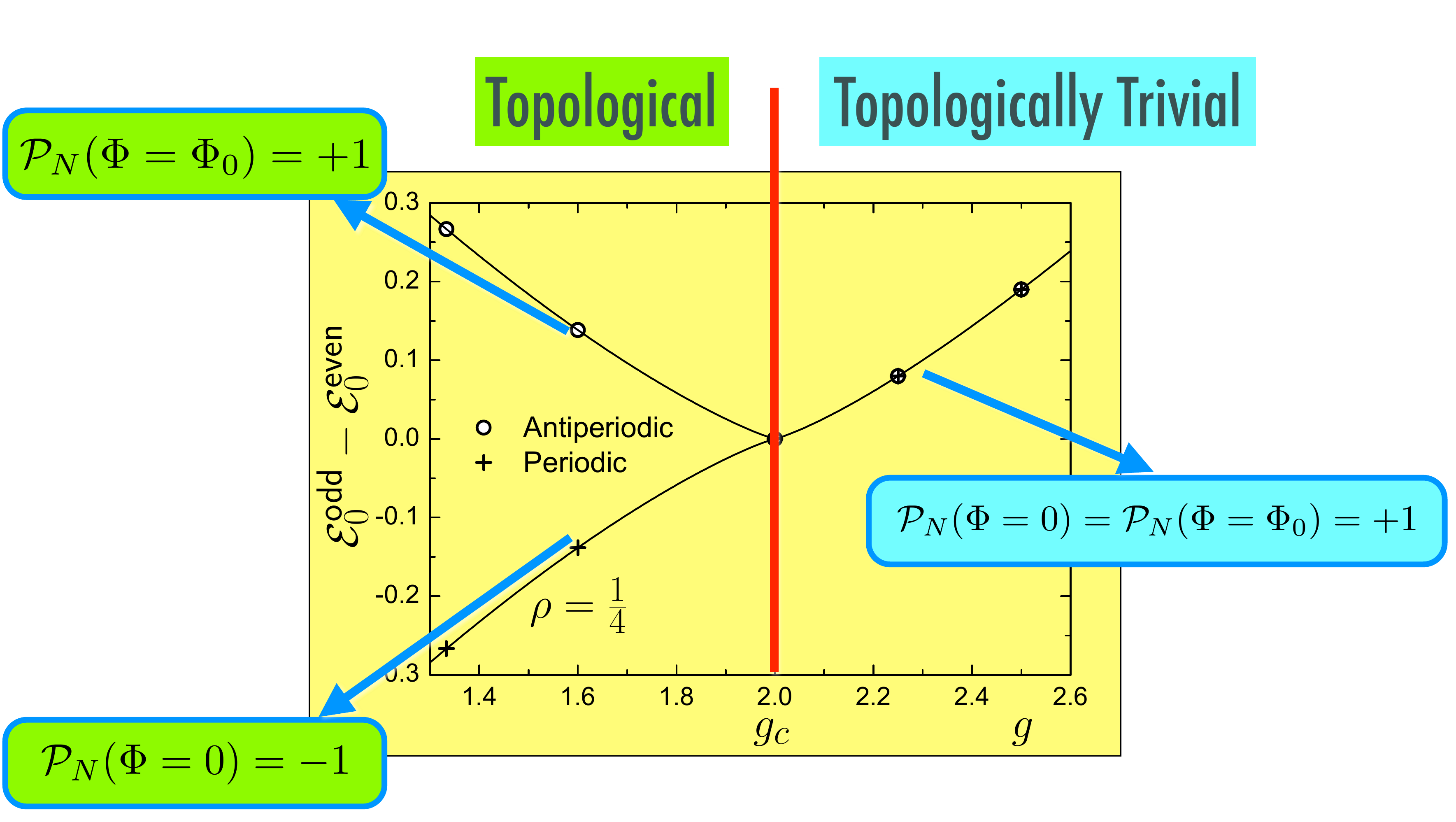}}
 \caption{Ground state energy differences (in units of $t_1\equiv 1$, for $t_2=0$) for even 
 ($N=2M)$ and odd ($N=2M\pm 1$) number of fermions, and  with 
 periodic ($\phi=0$) or antiperiodic ($\phi=2\pi$) boundary conditions. 
 The odd-even difference is shown as a function of the interaction strength $g$ for a finite 
 system (data points, for $N=512$, $L=2048$) and in the thermodynamic limit (continuous lines). 
 The topologically non-trivial state is entered for $g<g_c=2$. Also indicated are 
 the values of the fermion parity switches ${\cal P}_N(\Phi)$ across the transition. }
 \label{fig4}
\end{figure}

\begin{figure}[tb]
\centerline{\includegraphics[width=0.8\columnwidth]{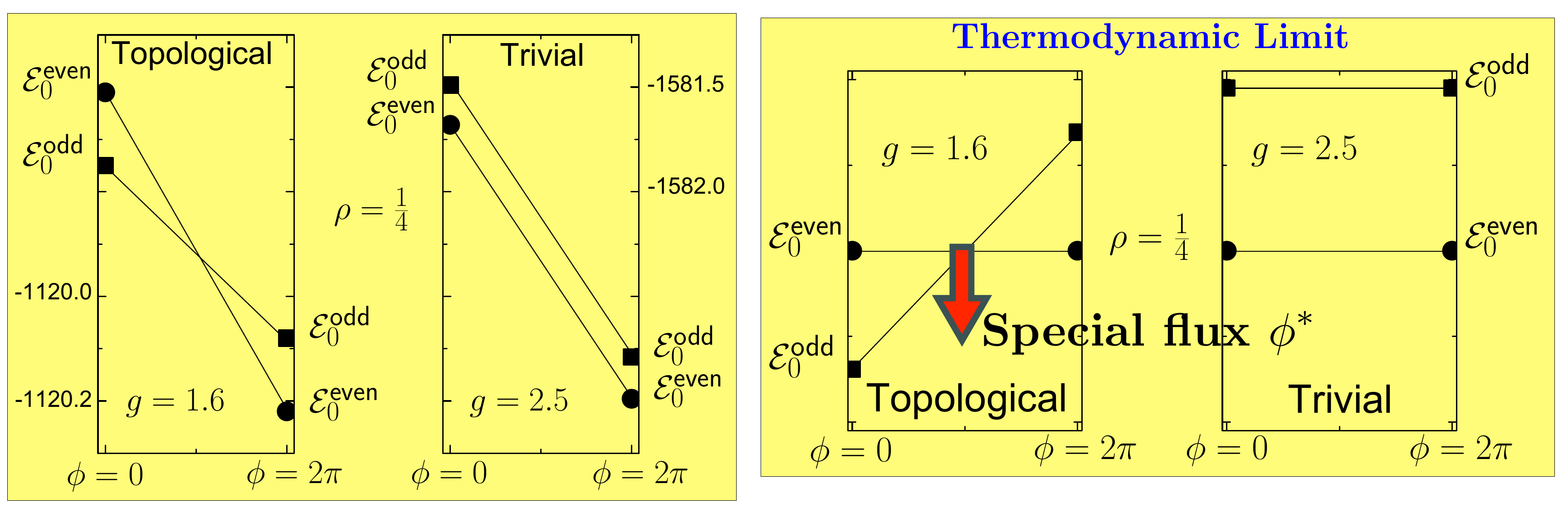}}
 \caption{Ground state energies for  even and odd number of fermions, 
for the finite system (left panel) and in the thermodynamic limit 
 (right panel), illustrating the fermion parity switches ${\cal P}_N(\phi)$. Notice 
 the special role played by the flux $\phi^*$ at the crossing (degeneracy) point. }
 \label{fig5}
\end{figure}

To identify the fermion parity switches we calculate the ground state 
energy ${\cal E}_0^\phi(N)$ for a given number $N$ of fermions in the 
chain of length $L$, with periodic ($\phi=0$) or antiperiodic 
($\phi=2\pi$) boundary conditions, and compare ${\cal E}_0^{\sf odd}(\phi)=
\tfrac{1}{2}{\cal E}_0^\phi(N+1)+\tfrac{1}{2}{\cal E}_0^\phi(N-1)$ 
and ${\cal E}_0^{\sf even}(\phi)={\cal E}_0^\phi(N)$, where we assumed 
$N$ even.  The difference (inverse compressibility) 
$\chi(\phi)= {\cal E}_0^{\sf odd}(\phi)-{\cal E}_0^{\sf even}(\phi)$ determines 
${\cal P}_N(\phi)={\rm sign}\,\chi(\phi)$, so it has the opposite sign at $\phi=0$ and $\phi=2\pi$ 
in the topologically non-trivial phase. We also find that ${\cal P}_{N\in {\sf even}}(\phi)=-{\cal P}_{N\in {\sf odd}}(\phi)$
in the topologically non-trivial phase. 
The results, shown in Figs. \ref{fig4} and \ref{fig5}, unambiguously demonstrate the topologically non-trivial 
nature of the superfluid for $g<g_c$ --- both in a finite system 
and in the thermodynamic limit, and without relying on any mean-field
approximation. 

The ground state of the odd ($2M\pm1$) system strongly depends on the boundary 
conditions. For periodic boundary conditions the unpaired particle always occupies the 
$k_0=0$ level, while for the antiperiodic case it starts blocking the Fermi 
momentum $k_F=k_0$ at $g=0$, continuously decreasing its modulus with increasing $g$, 
up to $k_0=\pi/L$ at $g_0\sim 1.1936$ ($\rho=1/4$), corresponding to $\mu = \Delta^2$ 
in the thermodynamic limit. 
In that limit $\chi(\phi)$ has a  particularly simple form: 
$\chi(0)=-8\mu$, and $\chi(2\pi)=8 |\mu|$ for $g>g_0$. 


\section{Many-body Majorana Zero-energy Modes}
\label{NCMajoranas}

The subject of this section is conceptually rather challenging. 
It is convenient to begin by recalling how  Majorana modes
emerge within the standard mean-field framework before we tackle
this problem in a many-body setting.

\subsection{Majorana modes in the mean-field approximation}
\label{mf}

In the mean-field description, i.e.,  the 
Bogoliubov-de Gennes ({\sf BdG}) approach, a second-quantized 
Hamiltonian describing a generic fermionic superfluid is \cite{MF,equivalence}
\begin{eqnarray}
{H}_{\sf mf}=\sum_{i,j=1}^L\Big[K_{ij}c_i^\dagger c^{\;}_j
+\frac{1}{2}\Delta^{\;}_{ij}c_i^\dagger c_j^\dagger
+\frac{1}{2}\Delta^*_{ij}c^{\;}_jc^{\;}_i\Big],
\label{Hfreep}
\end{eqnarray}
in terms of creation/annihilation operators $c^\dagger_j/c^{\;}_j$
of a fermion ($\{c^{\;}_i,c^\dagger_j\}=\delta_{ij}$) in the 
single-particle orbital $\phi_j$. The labels $i,j$ subsume arbitrary 
quantum numbers like position or momentum, band index, spin and orbital
angular momentum, and so on. The total number of single-particle orbitals 
is $L<\infty$ in the presence of infrarred (volume) and ultraviolet 
(lattice spacing) cutoffs. In Eq. \eqref{Hfreep}, the one-body kinetic
energy and mean-field pairing interaction matrices
\begin{eqnarray}
\mathbf{K}^\dagger=\mathbf{K},\quad \mathbf{\Delta}^T=-\mathbf{\Delta} ,
\end{eqnarray}
encode the relevant effective interactions of the superfluid system. Here 
$^\dagger$ is the adjoint, $^T$  the transpose, and $^*$ 
complex conjugation of a matrix.

In Nambu form, the Hamiltonian ${H}_{\sf mf}$ can be rewritten as
\begin{eqnarray}
{H}_{\sf mf}=\frac{1}{2} \, \hat{\phi}^\dagger \, H_{\sf BdG} \,  \hat{\phi} +\frac{1}{2}{\sf Tr}\,\mathbf{K} ,
\end{eqnarray}
where the Nambu column vector of fermion operators is given by
\begin{eqnarray} \hspace*{-0.5cm}
 \hat{\phi} = \binom{\hat c}{\hat c^\dagger} , \mbox{ with }  \hat{\phi}_j=c_j 
 \ , \  \hat{\phi}_{L+j}=c^\dagger_j \quad (j=1,\cdots, L).
\end{eqnarray}
Then the  {\sf BdG} single-particle Hamiltonian is
the $2L \times 2 L$ matrix  
\begin{eqnarray}\hspace*{-0.5cm}
H_{\sf BdG}&=&\begin{pmatrix}
\mathbf{K}& \mathbf{\Delta}\\
-\mathbf{\Delta}^*& -\mathbf{K}^*
\end{pmatrix} = 
\im\mathds{1}\otimes  \Im(\mathbf{K})+\im\tau^x\otimes \Im(\mathbf{\Delta})
+\im \tau^y\otimes \Re(\mathbf{\Delta})+\tau^z\otimes \Re(\mathbf{K})  ,
\end{eqnarray}
where $\tau^\nu$, $\nu=x,y,z$, are Pauli matrices, and 
$\Re(\cdot) (\Im(\cdot))$ denotes the real (imaginary) part of the matrix $\cdot$. 

No matter what the specific form of the matrices $\mathbf{K}$ and $\mathbf{\Delta}$ may be, the 
single-particle Hamiltonian $H_{\sf BdG}$ always anticommutes with 
the antiunitary  operator 
\begin{eqnarray}
\mathcal{C}=\mathcal{K}\tau^x\otimes \mathds{1},\quad \mathcal{C}^2=\mathds{1} ,
\end{eqnarray}
of particle-hole or charge-conjugation symmetry, $\{H_{\sf BdG},\mathcal{C}\}=0$.
Here \(\mathcal{K}\) denotes complex conjugation. It follows that 
the single-particle energy spectrum 
\begin{eqnarray}
H_{\sf BdG} \, \varphi_n&=& \epsilon_n \varphi_n \ ,  \ \ \  \varphi_n=\binom{X^{(n)}}{Y^{(n)}} \ ,  \ 
n=-L,-L+1,\cdots, -1,1, \cdots, L-1, L , 
\end{eqnarray}
is symmetrically distributed around zero and so it is automatically 
particle-hole symmetric, $\epsilon_n=-\epsilon_{-n}$. 
In turn, the single-particle spectrum can be used to write down $H_{\sf mf}$ 
\begin{eqnarray}
H_{\sf mf}=\sum_{n>0} \epsilon_n \, f_n^\dagger f^{\;}_n
+\frac{1}{2} \Big ( {\sf Tr}\,\mathbf{K} - \sum_{n>0} \epsilon_n \Big )
\end{eqnarray}
in terms  of quasi-particle operators
\begin{eqnarray}
f^\dagger_n&=& \sum_{j=1}^L (X^{(n)}_j \, c^\dagger_j + Y^{(n)}_j \, c^{\;}_j).
\end{eqnarray}

Assume now the existence of an {\it exact} zero energy mode, then it must be, at least, two-fold degenerate
\begin{eqnarray}
\epsilon_1=0=\epsilon_{-1} ,
\end{eqnarray}
and the associated single-particle states, $\varphi_1$, $\varphi_{-1}$, 
yield the quasi-particle operators
\begin{eqnarray}
f^\dagger_1= \sum_{j=1}^L X^{(1)}_j \, (c^\dagger_j +  c^{\;}_j) \ ,  \ X^{(1)}_j \in \mathds{R} \  \ 
, \ \ 
f^\dagger_{-1}= \sum_{j=1}^L X^{(-1)}_j \, (c^\dagger_j -  c^{\;}_j) \ ,  \ \im X^{(-1)}_j \in \mathds{R}.
\end{eqnarray}
These zero-energy quasi-particles are necessarily  Majorana fermions because
\begin{eqnarray}
f^\dagger_1=f^{\;}_1, f^\dagger_{-1}=f^{\;}_{-1}.
\end{eqnarray} 
In other words, in the  {\sf BdG} formalism, zero-energy modes are 
Majorana fermions by default. Typically, in a finite-size system, Majorana quasi-particles 
are not exact zero-energy modes but, strictly speaking, {\it emerge} as such only 
in the thermodynamic ($L\rightarrow \infty$) limit.

\subsection{Many-body Majorana modes beyond mean-field theory} 

In the {\sf BdG} approximation,  Majorana 
zero-energy modes of an open chain and 
the anomalous response to a flux insertion $\Phi$, in a ring configuration,  are two related signatures of topological 
fermion superfluidity. Since the response to a flux 
insertion remains a valid test in a number-conserving system, 
provided the correct parity switch is computed \cite{rgkchain}, 
we want to show a candidate for an emergent zero-energy mode at those special flux 
values $\Phi^*=\frac{\phi^*}{2\pi} \Phi_0$ where a level crossing occurs (see Fig. \ref{fig5}), and the 
fermionic superfluid is {\it gapped}.

To make this point more transparent, and before introducing the many-body version of the 
Majorana modes, let us start by setting the stage. For a given 
flux $\phi$, the coherent superposition state
\begin{eqnarray}
|\Psi_0^{\sf odd}\rangle=\frac{|\Psi_0^{N+1}\rangle + 
e^{{\sf i} \varphi}|\Psi_0^{N-1}\rangle}{\sqrt{2}} ,
\label{psiodd}
\end{eqnarray}
with $N \in$ even, is not a stationary state of the {\sf RGK} wire
(eigenstate of $H_{\sf RGK}$). Rather, it satisfies
\begin{eqnarray}
H_{\sf RGK}|\Psi_0^{\sf odd}\rangle={\cal E}_0^{\sf odd}(\phi)|\Psi_0^{\sf odd}\rangle+
\delta {\cal E}_0^{\sf odd}(\phi) |\widetilde{\Psi}_0^{\sf odd}\rangle ,
\end{eqnarray}
where 
\begin{eqnarray}
{\cal E}_0^{\sf odd}(\phi)&=&\frac{1}{2}({\cal E}_0^{\phi}(N+1)+
{\cal E}_0^{\phi}(N-1)), \ \ 
 \delta {\cal E}_0^{\sf odd}(\phi)=\frac{1}{2}
({\cal E}_0^{\phi}(N+1)-{\cal E}_0^{\phi}(N-1)) , \mbox{ and } \nonumber \\
|\widetilde{\Psi}_0^{\sf odd}\rangle&=&
\frac{|\Psi_0^{N+1}\rangle - e^{{\sf i} \varphi}|\Psi_0^{N-1}\rangle}{\sqrt{2}}.
\end{eqnarray}
The very important orthogonality relation 
\begin{eqnarray}
\langle {\Psi}_0^{\sf odd}|\widetilde{\Psi}_0^{\sf odd}\rangle= 0,
\end{eqnarray}
implies
\begin{eqnarray}
\langle\Psi_0^{\sf odd}|H_{\sf RGK}|\Psi_0^{\sf odd}\rangle=
{\cal E}_0^{\sf odd}(\phi).
\end{eqnarray}
On the other hand, the even parity state $|\Psi_0^{\sf even}\rangle=
|\Psi_0^{N}\rangle$ is a stationary state of $H_{\sf RGK}$
\begin{eqnarray}
H_{\sf RGK}|\Psi_0^{\sf even}\rangle=
{\cal E}_0^{\sf even}(\phi)|\Psi_0^{\sf even}\rangle,
\end{eqnarray}
where ${\cal E}_0^{\sf even}(\phi)={\cal E}_0^{\phi}(N)$.

Let \(\phi^*\) denote the value of the flux such that
\begin{equation}
{\cal E}_0^{\sf even}(\phi^*)={\cal E}_0^{\sf odd}(\phi^*).
\end{equation} 
Any continuous path in Hamiltonian space connecting our {\sf RGK} wire at
flux \(\phi=0, 2\pi\) will realize at least one such flux $\phi^*$. In 
particular, the {\sf RGK} chain, as defined in \(k\)-space, is exactly solvable 
for arbitrary values of \(\phi\) and  provides one possible interpolation.
However, the values \(\phi=0, 2\pi\) are special in that they make 
the position-representation of the  {\sf RGK} model physically appealing and  local. 
At such a $\phi^*$, by definition, the following relation is exactly satisfied
\begin{eqnarray}
\langle\Psi_0^{\sf even}|H_{\sf RGK}|\Psi_0^{\sf even}\rangle=
\langle\Psi_0^{\sf odd}|H_{\sf RGK}|\Psi_0^{\sf odd}\rangle .
\end{eqnarray}

We argued in Ref. \cite{rgkchain} that at $\phi^*$, one can define the 
zero-energy modes
\begin{eqnarray}
\Gamma_1=\hat{T}+\hat{T}^\dagger \ , \  \mbox{ and } {\sf i} \Gamma_2=\hat{T}-\hat{T}^\dagger  ,
\end{eqnarray}
in terms of the transition operators
\begin{eqnarray}
\hat{T}= |\Psi_0^{\sf even}\rangle \langle \Psi_0^{\sf odd}|, \  \ \hat{T}^2=0 , 
\ \ \{\hat{T},\hat{T}^\dagger\}=\hat{P}_0 ,  
\end{eqnarray}
with $\hat{P}_0= |\Psi_0^{\sf even}\rangle \langle \Psi_0^{\sf even}| +
|\Psi_0^{\sf odd}\rangle \langle \Psi_0^{\sf odd}|$ the projector onto the lowest-energy subspace.
This, in turn, implies
\begin{eqnarray}
\Gamma_1^2=\hat{P}_0= \Gamma_2^2 \ , \ \mbox{ and } 
\{\Gamma_1,\Gamma_2\}=0 ,
\end{eqnarray}
and, as we argue below, they are  natural candidates for emergent 
Majorana modes. We arrive at this conclusion after taking full advantage 
of the level crossing-like condition that we have established for a 
number-conserving topological superfluid (fermion parity switches),  
and a particular property that only happens at $\phi^*$, as we explain 
next.

Consider the commutators
\begin{eqnarray}
{[}H_{\sf RGK}, \Gamma_1]&=&\delta {\cal E}_0^{\sf odd}(\phi^*)( 
|\widetilde{\Psi}_0^{\sf odd}\rangle\langle\Psi_0^{\sf even}|-
|{\Psi}_0^{\sf even}\rangle\langle\widetilde{\Psi}_0^{\sf odd}|), \nonumber \\
{[}H_{\sf RGK}, \Gamma_2]&=&\!\!\!{\sf i} \ \delta {\cal E}_0^{\sf odd}(\phi^*)( 
|\widetilde{\Psi}_0^{\sf odd}\rangle\langle\Psi_0^{\sf even}|+
|{\Psi}_0^{\sf even}\rangle\langle\widetilde{\Psi}_0^{\sf odd}|) .
\end{eqnarray}
At this particular flux $\phi^*$, the energy difference 
$\delta {\cal E}_0^{\sf odd}(\phi^*)$ scales to zero in the infinite 
size (thermodynamic) limit  (at constant density $\rho=N/L$). We 
do not know the way it scales to zero but we speculate that it decays 
in an algebraic fashion as opposed to exponentially by comparison with 
the mean-field result, which can be determined quite accurately. 
The operators \(\Gamma_1, \Gamma_2 \equiv \Gamma_{1,2}\) are the 
simplest possible candidates for emergent zero modes that combine all 
of the available quantitative information.

Since the modes \(\Gamma_{1,2}\) would be zero modes of a ring configuration,
generated by a flux insertion, they are not expected to be localized. By analogy
to the basic Kitaev wire in a ring configuration, these Majorana modes generated
by a flux insertion are delocalized. On the other hand, had we used 
open boundary conditions, instead of a ring (flux) configuration, we would expect 
these modes to be in some appropriate sense localized on the edges of the wire \cite{example}; however, 
our {\sf RGK} wire is neither exactly solvable in this case nor it is
clear how the exact solvability condition for the other boundary conditions 
could help to determine localization.  It is important to realize that $\Gamma_{1,2}$ connect the 
even and odd parity sectors; their actions, for instance, on the 
coherent ground state $|\Psi_0^{\sf even}\rangle$ are
\begin{eqnarray}
\Gamma_1 |{\Psi}_0^{\sf even}\rangle = |{\Psi}_0^{\sf odd}\rangle \ , \ \ 
\Gamma_2 |{\Psi}_0^{\sf even}\rangle = {\sf i} |{\Psi}_0^{\sf odd}\rangle , 
\end{eqnarray}
showing that the Majorana operators can realize states that seem to violate 
a charge superselection rule.  
The parity operator, on the other hand, 
\begin{eqnarray}
\hat{\cal P}  =  \im \Gamma_2\Gamma_1=
|\Psi_0^{\sf even}\rangle \langle \Psi_0^{\sf even}| -|\Psi_0^{\sf odd}\rangle \langle \Psi_0^{\sf odd}| \ , \ 
\hat{\cal P} |\Psi_0^{\sf even, odd}\rangle = \pm  |\Psi_0^{\sf even, odd}\rangle ,
\label{parityexp}
\end{eqnarray}
does not connect the odd and even sectors.

Notice that the algebra of the \(\Gamma_{1,2}\)
zero modes is not the main issue at stake, since such an algebra 
could be obtained in other systems with ground-state degeneracies. However, there are additional 
reasons to  claim that the interacting \(\Gamma_{1,2}\) modes 
are the modes that evolve into the standard Majorana modes in the 
mean-field approximation. First, there is the role of flux insertion in 
the construction of the modes, characteristic of topological fermion superfluidity. 
Second, the specific form of the \(\hat{T}\) operator immediately shows that the
\(\Gamma_{1,2}\) modes anticommute with fermionic parity, and this is one of 
the key properties of  Majorana modes. Lastly, the modes rely on  
{\it coherent} superpositions of states with a different number of particles, which is also 
unique to Majorana physics. 

In summary, the operators $\Gamma_{1,2}$ suggest a simple way in which zero 
energy modes, which, in addition, are non-number-conserving, might 
emerge in a number-conserving topological fermionic superfluid. Their emergence 
raises an important question: Is it really possible to prepare 
and manipulate experimentally coherent superpositions of states
with different particle numbers such as $|\Psi_0^{\sf odd}\rangle$?  
This is of utmost importance, since if it is not possible, because 
of superselection rules, a message one could get 
from our number conserving analysis is that it is not viable to 
physically manipulate Majorana modes for quantum information purposes.

\section{Preparation and Manipulation of Majoranas:
The Superselection Viewpoint}
\label{SSR}

Topologically non-trivial $p$-wave superconducting chains, that is, 
systems modeled by the Majorana chain of Kitaev or the {\sf RGK} wire 
of this paper for example, are systems of great interest in experimental 
quantum information processing. Their ground-state subspace, being 
two-fold degenerate and separated from the quasi-particle continuum by 
an energy gap, has the potential to encode a single qubit.  
The natural basis of this subspace would consist of the state with 
no Bogoliubov quasi-particles (the vacuum of the system) and a second state obtained 
by applying a (generalized) Majorana mode localized on either end of the 
chain to this vacuum. Since these states differ in their fermionic parity, 
a symmetry that cannot be explicitly broken, at least the degeneracy of 
this subspace is expected to be resilient to decoherence by the environment. 
Finally, the operators that connect these states, the Majorana modes, 
are (presumably, for the {\sf RGK} chain) localized in position space \cite{example}. Hence, 
the setup seems good and ripe for quantum manipulation and control. What 
is it to be learned from incorporating particle-number conservation into 
the picture? 

Models like the {\sf RGK} chain demonstrate the consistency between 
the requirements of topological non-triviality and a ground-state subspace 
with a basis that has a definite fermionic number (as opposed to just 
definite fermionic parity): The vacuum has some fixed number of fermions, 
and the ``topologically degenerate" state has one more or less fermion. 
This result is encouraging since it shows that fermionic 
parity switches, as a criterion for topological fermion superfluidity, survive 
beyond the mean-field approximation. But it also shows this: 
\vspace*{0.2cm}

\noindent \fbox{
\parbox{15.5cm}{\bf  Encoding a qubit in the ground-state subspace 
of a $p$-wave superconducting wire, and its control,  
require that {\it particular coherent} superpositions of states differing in particle  
number should be physically/experimentally realizable.} }

\vspace*{0.2cm}
There are however good reasons, and even mathematical proofs sometimes,
for believing such coherent superpositions do not in fact represent 
physical states. This is the subject of superselection rules.

Let us begin by recalling the best established aspects of 
superselection rules. The traditional pieces of the subject 
are lucidly described in Ref.\,\cite{wightman95}, and 
recent views are summarized in Ref.\,\cite{bartlett07}.
In the standard approach to superselection rules, the systems 
under consideration are taken to conserve 
the physical quantity that is argued to be superselected.

Consider for concreteness a system of spin-1/2 fermions 
\begin{eqnarray}
\{c^{\;}_{i,\sigma},c^{\;}_{j,\sigma'}\}=0,\quad 
\{c^{\;}_{i,\sigma},c_{j,\sigma'}^\dagger\}=\delta_{ij}\delta_{\sigma\sigma'},
\end{eqnarray}
with total spin angular momentum operators
\begin{eqnarray}
{\sf S}^\nu=\frac{1}{2}
\sum_j c_{j,\sigma}^\dagger \, (\tau^\nu)_{\sigma\sigma'} \, c_{j,\sigma'}
\quad (\nu=x,y,z) .
\end{eqnarray}
Since the unitary map $e^{\im \, \theta \, {\sf S}^z}$ induces the following 
transformation
\begin{eqnarray}
e^{\im \, \theta \, {\sf S}^z}c_{j,\sigma}^\dagger e^{-\im \, \theta \, {\sf S}^z}
=e^{\im \, \theta/2} \, c_{j,\sigma}^\dagger,
\end{eqnarray}
the (Slater determinant) basis of many-body states 
\begin{eqnarray}
c_{j_1,\sigma_1}^\dagger \dots c_{j_N,\sigma_N}^\dagger|0\rangle
\end{eqnarray}
(\(|0\rangle\) is the Fock vacuum) is such that 
states with \(N\) even are unchanged by a $2 \pi$ rotation, while 
 states with \(N\) odd are changed by a $e^{\im \pi}=(-1)$ factor. 

Imagine now trying to prepare a coherent superposition like 
\begin{eqnarray}
\alpha_1 |0\rangle+\alpha_2 \, c_{j,\sigma}^\dagger|0\rangle \quad 
\quad (\alpha_{1,2} \in \mathbb{C}) .
\end{eqnarray} 
The best established superselection rule, known as the boson/fermion 
or univalence superselection rule \cite{wightman95}, states that such 
coherent superpositions are not physical 
or physically realizable. 
While results such as  
\begin{eqnarray}
\alpha_1 |0\rangle+\alpha_2 \, c_{j,\sigma}^\dagger|0\rangle
\ \stackrel{e^{\im \, 2\pi \, {\sf S}^z}}{\longrightarrow}\ 
\alpha_1 |0\rangle-\alpha_2 \, c_{j,\sigma}^\dagger|0\rangle
\end{eqnarray} 
are not quite enough to completely prove the superselection rule,
an appeal to the time-reversal operation  
$\hat{\Theta}=U \, {\cal K}$ 
(with $U$ a unitary operator associated to the spin-1/2 of the 
fermions),  and its resulting defining relation
\begin{eqnarray}\label{fpt}
\hat{\Theta}^2=(-1)^{2{\cal J}}=(-1)^{\hat{N}},
\end{eqnarray} 
with \({\cal J}\) the spin of the representation of rotations induced by
\({\sf S}^\nu\) on states \(|\Psi_N\rangle\) with \(N\) and 
only \(N\) fermions, i.e., \(\hat N|\Psi_N\rangle=N |\Psi_N\rangle\), 
completes the proof \cite{wightman95}. Incidentally, Eq.\,\eqref{fpt}
explains why fermionic parity cannot be explicitly broken in systems 
of fermions. 

Let us rephrase these arguments in the context of the Majorana modes
of the {\sf RGK} chain. Two consecutive applications of the time-reversal 
operation $\hat{\Theta}$ should leave any physical ray  unchanged.  
Consider the coherent superpositions 
\begin{eqnarray}
|\Psi_\pm \rangle= \frac{|\Psi_0^{\sf even}\rangle\pm e^{\im \vartheta}|\Psi_0^{\sf odd}\rangle}{\sqrt{2}} .
\end{eqnarray}
If one applies time-reversal twice to $|\Psi_\pm \rangle$ it implies that 
\begin{eqnarray}
\hat{\Theta}^2 |\Psi_\pm \rangle= |\Psi_\mp\rangle  ,
\end{eqnarray}
therefore,  not  leaving the ray $|\Psi_\pm \rangle$ invariant. 
Consequently, the Wick-Wightman-Wigner boson/fermion superselection 
rule \cite{wick52}, applied to the coherent 
superposition $|\Psi_\pm \rangle$ indicates that there is no physical observable that can 
measure the relative phase $e^{\im \vartheta}$ between the even (bosonic) and odd (fermionic) sectors. Then, 
Hermitian operators connecting those sectors, such as  $\Gamma_{1,2}$, are not measurable.  
Since parity $\hat{\cal P}$ does not connect the even and odd sectors it could, in principle, be 
measurable. However, notice from Eq. \eqref{parityexp} that parity $\hat{\cal P}$ requires 
preparation of the coherent superposition $|\Psi_0^{\sf odd}\rangle$, Eq. \eqref{psiodd}.

\section{Conclusions and Outlook}
\label{outlook}


In this paper we presented a criterion for characterizing 
interacting, particle-number conserving fermionic superfluids
as topologically trivial or non-trivial \cite{rgkchain}. Our 
criterion combines the many-body ground-state energies of systems with 
$N$, $N-1$ and $N+1$ fermions, computed for both periodic and anti-periodic 
boundary conditions, into an observable with the interpretation
of an inverse compressibility that keeps track of switches in 
fermionic parity as the phase \(\phi\) specifying twisted boundary 
conditions evolves from \(\phi=0\) (periodic boundary conditions) to 
\(\phi=4\pi\), passing through \(\phi=2\pi\) (antiperiodic boundary 
conditions). The behavior of this inverse compressibility in the 
thermodynamic limit dictates the topologically trivial or non-trivial 
character of the superfluid phase in question.  The experimental 
signature, directly related to this test  of topological non-triviality 
beyond mean-field, is the $4\pi$-periodic Josephson effect.\cite{rgkchain}

Realizing an exactly-solvable, particle-number conserving, minimal
generalization of the Majorana chain of Kitaev has helped enormously to
clarifying what is entailed by the notion of a topological superfluid
beyond the mean-field approximation. In particular, because of
its Bethe ansatz solution, we were able to investigate 
our criterion for topological non-triviality very explicitly 
in the Richardson-Gaudin-Kitaev ({\sf RGK}) wire. Notice that
``minimal generalization" implies that the number-conserving 
wire should be gapped inside the various superfluid phases. 
This explains why the pairing interaction in the {\sf RGK} chain must be long-range,
while the complete integrability condition merely tweaks the particular
shape of this interaction. By the Mermin-Wagner (or Coleman in field 
theory) theorem, compact, continuous symmetries cannot be spontaneously 
broken in one dimension, for systems with finite-range interactions. Our
{\sf RGK} wire escapes this general result by violating this last
condition, and so manages to display a true gap in the thermodynamic
limit associated to the spontaneous, but not explicit as in mean-field approaches,
breaking of particle-number conservation and the presence of long-range order. 

On general grounds, one expects that fermionic parity switches 
should be affected by zero-energy modes of Majorana character. 
Stimulated by our findings we introduced in Section \ref{NCMajoranas}, and provided a  
physical meaning to, the concept of many-body Majorana zero-energy modes. 
We believe we have managed to reasonably 
show that our zero-energy modes, beyond mean-field, should indeed be 
thought of as Majorana modes. Three are the reasons behind our claim. First, even 
though they emerge in a particle-number conserving system, our many-body 
Majorana modes do not conserve
particle number. Second, they satisfy the Majorana algebra and,
perhaps even more importantly, they precisely anticommute with 
fermionic parity. And third, it is possible to argue, partly 
because they anticommute with fermionic parity, that our many-body
Majorana modes should evolve into the corresponding mean-field 
modes of Section\,\ref{mf}.

Let us now discuss the issue of quantum manipulation and control 
of Majorana modes.
It is convenient to recall briefly the importance of the Majorana 
chain from the viewpoint of quantum information processing. 
One could, in principle, encode a single qubit  since the 
ground-state energy level is 
 two-fold degenerate. However, 
 this is not sufficient. What is 
important is that this subspace is resilient to errors characterized by 
quasi-local operators quadratic in fermions, and those errors preserve fermionic parity. 
On the other hand, the ground-state degeneracy of the Majorana 
chain with open boundaries is associated to the spontaneous breaking 
of fermionic parity. 
In general, knowing which symmetry is responsible for the degeneracy 
makes it simple to lift: just break  the 
responsible symmetry explicitly. However, this is not simple for fermionic
parity. Quasi-local fermionic operators that fail to commute 
with fermionic parity, must also necessarily fail to commute 
with each other at long distances (think for example of the
creation and annihilation operators in real space, and products
of three creation and annihilation operators, etc.). Hence, because 
of the interplay between locality and fermionic statistics, it is not 
possible to break fermonic parity explicitly for a closed system.  

As explained in Section\,\ref{SSR} we believe that 
the traditional theory of superselection rules enforces restrictive 
constrains to the idea that zero-energy modes of fermion superfluids 
anticommuting with fermionic parity may be quantum manipulated. The 
problem in short is that fermionic parity is a superselected symmetry,
for at least two reasons. First, there is the relations between
fermionic parity and the square of the physical time-reversal operation.
Second, there is the relation between fermionic parity and electric
charge, a quantity that is superselected due to Gauss law \cite{wightman95}.
Incidentally, the fact the fermionic parity is superselected in a 
closed system explains from a different point of view why
 it cannot be explicitly broken. 
  
The traditional view on superselection rules has, however, been
challenged from time to time. Aharonov and Susskind \cite{aharonov67} 
were the first to emphatically assert that there are no superselection
rules in non-relativistic quantum mechanics. Their argument, adapted to 
our particle-conserving topological fermion superfluid, requires the 
introduction of a  {\it suitable designed environment} that couples 
to our system and acts as a reference system. In that way, they would 
claim that it is possible to measure the phase $e^{\im \vartheta}$ 
relative to that environment represented by a reference state. 
In other words, by suitably coupling our {\sf RGK} wire to a reservoir that allows 
for interchange of fermions, simultaneously violating the conservation of fermionic parity, 
one could potentially generate a many-body Majorana zero-energy mode.


Regardless, we would like to point out that it will be, {\it at least}, 
extremely hard to achieve such a level of coherence experimentally in, for instance, 
solid state nanowires. The
computational power of Majorana braiding is rooted in its non-Abelian, dynamically 
generated,  
properties that follow simply from its algebra and the existence of an excitation 
energy gap.


\section*{Acknowledgements}

We acknowledge illuminating discussions with Tony Leggett, Manny Knill, and Babak Seradjeh.

\end{document}